\documentclass[aps,onecolumn,nofootinbib,amsmath,amssymb,groupedaddress]{revtex4}
\usepackage{epsfig}
\usepackage{graphicx}

\usepackage{subcaption}

\usepackage{bm,dsfont,braket,mathrsfs}
\usepackage{color}

\newcommand{\be}{\begin{eqnarray}}
\newcommand{\ee}{\end{eqnarray}}

\newcommand{\rv}[1]{{#1}}

\begin{document}

\title{Quantum Hall conductivity in the presence of interactions }

\author{Xi Wu\footnote{Electronic address: wuxi5949@gmail.com}}
\affiliation{Physics Department, Ariel University, Ariel 40700, Israel}

\author{M.A.~Zubkov\footnote{Electronic address: mikhailzu@ariel.ac.il, on leave of absence from NRC "Kurchatov Institute" - ITEP, B. Cheremushkinskaya 25, Moscow, 117259, Russia} }
\affiliation{Physics Department, Ariel University, Ariel 40700, Israel}

\date{\today}

\begin{abstract}
We discuss quantum Hall effect in the presence of arbitrary pair interactions between electrons. It is shown that irrespective of the interaction strength the Hall conductivity is given by the filling fraction of Landau levels averaged over the ground state of the system. This conclusion remains valid for both integer and fractional quantum Hall effect.
\end{abstract}
\pacs{
}

\maketitle

%



\section{Introduction}

Integer quantum Hall effect (IQHE) in the presence of constant external magnetic field is quantized, and its conductivity is an integer multiple of  $e^{2}/h$ \cite{Klitzing}. Although the systems that possess IQHE may be extremely complicated and may include interactions and disorder, the quantization of Hall conductivity is precise. It is in fact so precise, that the present measurements of the fine structure constant are based on it. Without interactions the IQHE conductivity is a topological quantity expressed through the Berry curvature integrated over the occupied energy levels \cite{AS2,TKNN}. The presence of impurities does not alter this statement \cite{Bellissard, AS2index, AG}. The absence of corrections to the IQHE due to Coulomb interactions and impurities (in the presence of constant magnetic field) has been discussed widely (see, for example,  \cite{Alt0,Alt1}). In the presence of the variations of magnetic field (but without interactions) the IQHE conductivity is expressed through the topological invariant in \cite{ZW2019}. It has been shown that weak interactions do not alter this property as long as turning on interactions does not drive the system to a topological phase transition \cite{ZZ20192}. The situation in the presence of sufficiently strong interactions remains unclear, especially for the systems with fractional quantum Hall effect (FQHE). At the present moment the topological nature of FQHE remains an open question \cite{OpenProblems}. There were several attempts to explain the topological nature of both IQHE and FQHE in the presence of interactions in the framework of various phenomenological effective theories \cite{BFr,W, FrK, FZ, Z, FrS, FrST}. However, their direct connection was not established to the particular microscopic models.

In \cite{HM} the quantization of Hall conductivity has been proven mathematically for a particular lattice model. This proof presumably may be extended to the other systems with the non - degenerate ground states with a gap. The latter condition remains nontrivial
\cite{DFF, DFFRB,BHM,BH,MZ,BNY}, and cannot be proven analytically for the majority of interacting systems. In \cite{QHE_proof} the proof has been given that the Hall conductivity is quantized for the gapped interacting system with weak short - ranged interactions. This proof may be used, in particular, for the  Hofstadter \cite{AEG,Hof}
and Haldane \cite{H1} models with interactions. Notice, that the interacting Haldane model has been investigated numerically \cite{R2,R1}.
 Unfortunately, these results cannot be applied directly to the case when the most interesting long - ranged Coulomb interactions are present.
In \cite{IM,CH} relation of the quantum Hall effect (QHE) to Ward identities was discussed. The lattice version of Ward identities \cite{IM,CH} was used for the resummation of Feynman diagrams in \cite{QHE_proof}. A similar resummation technique was applied also in \cite{ZZ20192,ZZ20193} and earlier in \cite{BGPS, BMdensity, BMchiral, BM, BFM0, BFM1, GMP1,GMP2,Mbook}.

In the present paper we consider the QHE in the presence of constant magnetic field for the systems  without disorder with arbitrarily strong pair inter - electron interactions. In such systems this is not possible to speak about the completely occupied one - particle states. Interactions may, in principle, lead to the fluctuations of the occupation numbers even at zero temperature. We introduce operator of the filling fraction $\hat{\boldsymbol{\nu}} = \frac{\hat{\mathbb {N}}}{M}$, where $\hat{\mathbb {N}}$ is the operator of the number of electrons, while $M$ is the degeneracy of each Landau level (it does not depend on its number). We demonstrate (presumably, for the first time for such systems) that the Hall conductivity is given by the expectation value of $\hat{\boldsymbol{\nu}}$ in the ground state of the given system times the Klitzing constant  $e^{2}/h$. Being derived for the arbitrary pair interactions our results are valid for both IQHE and FQHE. At this stage we do not yet prove the topological invariance of the Hall conductivity in such systems (i.e. its robustness with respect to arbitrary variation of the system).

\section{Fixed numbers of electrons}
\label{Stand}
In this section we consider the interacting system with the constant homogeneous external magnetic field, and fixed number of electrons. The multi - particle Hamiltonian with the interaction term has the form:
\begin{eqnarray}\label{HCP}
	\hat{\cal {H}}&=& \sum_{a=1}^{N} \hat{\cal H}_0(-i\partial_{x^{(a)}} + e\mathbf{A}(x^{(a)}))+\frac{1}{2}\sum_{a \ne b}^{N} V(x^{(a)}-x^{(b)})\,,
\end{eqnarray}
In the following, we take a shorthand notation $\sum_a$ for $\sum_{a=1}^{N}$.
The Kubo formula for Hall conductance is
\be
	\sigma_{xy}&=&\frac{i\hbar} {\cal S}\sum_{k\ne 0}\frac{\langle 0 | \hat{J}_y |k\rangle \langle k|\hat{J}_x| 0 \rangle-\langle 0 | \hat{J}_x |k\rangle \langle k| \hat{J}_y | 0 \rangle}{({\cal E}_k-{\cal E}_{0})^2}\,.
\ee
where $\cal S$ is the area of the system. The total electric current may be written as
\be
\hat{J}_i=\frac{ie}{\hbar}[\hat{\cal H},\sum_a\hat{x}^{(a)}_i]\,,
\ee
Thus the Hall conductance of this quantum - mechanical system may be  written as
\be
	\sigma_{xy}&=&\frac{1}{i {\cal S}}\frac{e^2}{\hbar}\sum_{k\ne0}\frac{\langle 0 | [\hat{\cal H},\sum_b\hat{y}^{(b)}] |k\rangle\langle k| [\hat{\cal H},\sum_a\hat{x}^{(a)}] | 0\rangle-\langle 0 |  [\hat{\cal H},\sum_a\hat{x}^{(a)}] |k\rangle\langle k| [\hat{\cal H},\sum_b\hat{y}^{(b)}] | 0\rangle}{({\cal E}_k-{\cal E}_0)^2}\,.\nonumber
\ee
Written in a compact way, it becomes
\begin{eqnarray}
	\sigma_{xy} &=& -\frac{1}{i {\cal S}}\frac{e^2}{\hbar}\,\sum_{k\ne 0}   \, \epsilon_{ij}\,\Big[  \frac{1}{({\cal E}_k - {\cal E}_0)^2}  \langle 0| [\hat{\cal H}, \sum_a{\hat x}_i^{(a)}] | k \rangle    \langle k | [\hat{\cal H}, \sum_b{\hat x}_j^{(b)}] | 0 \rangle  \Big]_{A=0}\,, \label{sigmaHHI__}
\end{eqnarray}
where $| k \rangle$ labels a many-body state and $| 0 \rangle$ is the ground state, and $i,j= x, y$.

For the minimally coupled single-electron Hamiltonian $\hat{\cal H}_0(\boldsymbol{\pi})=\hat{\cal H}_0(\mathbf{p}+e\mathbf{A}) = \hat{\cal H}_0(-i\partial_{x} + e\mathbf{A}(x))$ we decompose the coordinates $x_1=x, x_2=y $ as follows:
$$
\hat{x}= \frac{\hat{\pi}_y}{e B} + \hat{X} = \hat{\xi}_x + \hat{X},$$ $$ \hat{y}= -\frac{\hat{\pi}_x}{e B} + \hat{Y}= \hat{\xi}_y + \hat{Y}\,,
$$
where $B=B_z=\partial_x A_y-\partial_yA_x$.
The gauge-independent commutation relations follow:
\begin{eqnarray}\label{xixxiy}
	[\hat{\xi}_x,\hat{\xi}_y] = -\frac{i\hbar}{e B}\,, \quad [\hat{X},\hat{Y}] =  \frac{i\hbar}{e B}\,,
\end{eqnarray}
\begin{eqnarray}\label{HXY}
	[\hat{\cal H}_0, \hat{X}] =  [\hat{\cal H}_0, \hat{Y}] =  0\,.
\end{eqnarray}

And  for the interacting Hamiltonian the commutation relations Eq. (\ref{xixxiy}), Eq. (\ref{HXY}) are generalized into
\begin{eqnarray}\label{SXI}
	[\sum_a\hat{\xi}^{(a)}_x,\sum_b\hat{\xi}^{(b)}_y] = -\frac{iN\hbar}{e B}\,, \quad [\sum_a\hat{X}^{(a)},\sum_b\hat{Y}^{(b)}] =  \frac{iN\hbar}{e B}\,,
\end{eqnarray}
\begin{align}\label{HSXY}
	[\hat{\cal H},\sum_a \hat{X}^{(a)}] =0 \,, \quad
	[\hat{\cal H},\sum_a \hat{Y}^{(a)}] =0 \,,
\end{align}

Substituting Eq.(\ref{SXI}) and  Eq.(\ref{HSXY}) into  Eq.(\ref{sigmaHHI__}), we get:
\begin{eqnarray}\label{NBS}
	\sigma_{xy}&=&   -\frac{1}{i {\cal S}}\frac{e^2}{\hbar}\,\sum_{k\ne 0}   \,\Big[  \frac{\epsilon_{ij}}{({\cal E}_k - {\cal E}_n)^2}  \langle 0| [\hat{\cal H}, \sum_a\hat{\xi}^{(a)}_i | k \rangle    \langle k | [\hat{\cal H}, \sum_a\hat{\xi}^{(a)}_j] | 0 \rangle  \Big]_{A=0} \,
\nonumber\\
	&=& \frac{1}{i {\cal S}}\frac{e^2}{\hbar}\,\sum_{k\ne 0}   \, \epsilon_{ij}\,\Big[  \langle 0| \sum_a\hat{\xi}^{(a)}_i | k \rangle    \langle k |  \sum_a\hat{\xi}^{(a)}_j | 0 \rangle  \Big]_{A=0}
\nonumber\\
	&=&  \frac{1}{i {\cal S}}\frac{e^2}{\hbar}  \,\Big[  \langle 0|  [\sum_a\hat{\xi}^{(a)}_x,\sum_b\hat{\xi}^{(b)}_y ]| 0 \rangle  \Big]_{A=0}
\nonumber\\
	&=&  - \frac{eN}{B {\cal S}} \,.
\end{eqnarray}
If we denote by $M$ the number of electron states in a fully occupied Landau level, then
\begin{eqnarray}
	|{B}|{\cal S}=M\Phi_0=\frac{2\pi \hbar}{|e|} M\,,
\end{eqnarray}
where
$\Phi_0$ is flux quanta. Then we have
\begin{eqnarray}\label{HCIF}
	\sigma_{xy}&=&\frac{Ne^2}{M h}\, {\rm sign}({B})\nonumber
	\\
	&=&\nu \, \frac{e^2}{h}{\rm sign}({B})\,,
\end{eqnarray}
where $\nu$ is the filling fraction. As a result, we see that only the filling fraction is relevant for the Hall conductivity for both integer and fractional quantum Hall effect, which has been observed in experiment. Coulomb interaction does not affect the steps up to Eq.(\ref{NBS}), and that is why (at least, in the absence of impurities) the Hall conductivity of integer quantum Hall effect can be calculated in the free electron system.

\section{The system described by a chemical potential}

Now we consider the situation, when the number of electrons is allowed to fluctuate, but  the chemical potential is introduced. The Hamiltonian will be written in the second-quantized form:
\begin{eqnarray}\label{HCP}
	\hat{\mathbb {H}}&=&\int d^2x^{}a^{\dagger}(x)(\hat{\cal H}_0-\mu)a(x)
	\nonumber\\
	&&+\frac{1}{2}\int d^2x \int d^2x'a^{\dagger}(x)a^{\dagger}(x')V(x-x')a(x')a(x)\,,
\end{eqnarray}
where $\hat{\cal H}^{}_0=\hat{\cal H}^{}_0(\mathbf{p}+e\mathbf{A})$ is a single-electron Hamiltonian minimally coupled to the background gauge field, while $a^{\dagger}(x^{})$ and $a(x^{})$ are the fermionic creation and annihilation operators in coordinate space.
Let us define the two single-body operators:
\begin{eqnarray}
	\hat{\mathbb {F}}:=\int d^2x  \,a^{\dagger}(x )\hat{F}(x )a(x )\,,\quad\quad
	\hat{\mathbb {G}}:=\int d^2x  \,a^{\dagger}(x )\hat{G}(x )a(x )\,,
\end{eqnarray}
Here operators $\hat{F}$ and $\hat{G}$ act on $a$ considered as a function of $x$. We omit possible internal symmetry indices for brevity. Formally the further expressions are valid for the spinless electrons, but in fact we can extend easily our consideration to the case, when the operators $a$ have indices, and operators $\hat{F}, \hat{G}$ act on those indices as well. In a similar manner we construct the two-body operator:
\begin{eqnarray}
	\hat{\mathbb {K}}:=\int d^2x'\int d^2x''  \,a^{\dagger}(x')a^{\dagger}(x'')\hat{K}(x',x'')a(x'')a(x')\,,
\end{eqnarray}
We have (see the derivation in Appendix \ref{Commutators in second quantization})
\begin{eqnarray}\label{FG}
	[\hat{\mathbb {F}}, \hat{\mathbb {G}}]&=&\int d^2x\int d^2x'\,[a^{\dagger}(x )\hat{F}(x )a(x ),a^{\dagger}(x')\hat{G}(x')a(x')]\nonumber\\
	&=&\int d^2x\,a^{\dagger}(x )[\hat{F}(x ),\hat{G}(x)]a(x)
	\,,
\end{eqnarray}
and
\begin{eqnarray}\label{FK}
	[\hat{\mathbb {F}}, \hat{\mathbb {K}}]&=&\int d^2x\int d^2x'\int d^2x''\,[a^{\dagger}(x )\hat{F}(x )a(x ),a^{\dagger}(x')a^{\dagger}(x'')\hat{K}(x',x'')a(x'')a(x')]\nonumber\\
	&=&\int d^2x\int d^2x'\,a^{\dagger}(x )a^{\dagger}(x')[\hat{F}(x ),\hat{K}(x,x')+\hat{K}(x',x)]a(x')a(x)
	\,.
\end{eqnarray}
Next, we generalize the quantum-mechanical coordinate-sum operators into the second-quantized coordinate-sum operators:
\begin{eqnarray}
	\sum\hat{\mathbf{x}}:=\int d^2x  \,a^{\dagger}(x ) \,\hat{x} \,a(x )\,,\quad\quad
	\sum\hat{\mathbf{y}}:=\int d^2x  \,a^{\dagger}(x ) \,\hat{y} \,a(x )\,,
\end{eqnarray}
\begin{eqnarray}
	\sum\hat{\boldsymbol{\xi}}_x:=\int d^2x  \,a^{\dagger}(x ) \,\hat{\xi}_x \,a(x )\,,\quad\quad
	\sum\hat{\boldsymbol{\xi}}_y:=\int d^2x  \,a^{\dagger}(x ) \,\hat{\xi}_y \,a(x )\,,
\end{eqnarray}
\begin{eqnarray}
	\sum\hat{\mathbf {X}}:=\sum\hat{\mathbf{x}}-\sum\hat{\boldsymbol{\xi}}_x\,,\quad\quad
	\sum\hat{\mathbf {Y}}:=\sum\hat{\mathbf{y}}-\sum\hat{\boldsymbol{\xi}}_y\,,
\end{eqnarray}
With Eq.(\ref{FG}) and Eq.(\ref{FK}), we can generalize expressions for the commutators from the last section to the present case (with constant chemical potential instead of the fixed number of particles).
\begin{eqnarray}\label{sxixy}
	[\sum\hat{\boldsymbol{\xi}}_x, \sum\hat{\boldsymbol{\xi}}_y]=-\frac{i\hbar}{e B}\int d^2x \,a^{\dagger}(x)a(x)\,,
	\quad\quad
	[\sum\hat{\mathbf {X}}, \sum\hat{\mathbf {Y}}]=\frac{i\hbar}{e B}\int d^2x \,a^{\dagger}(x )a(x )\,,
\end{eqnarray}
\begin{eqnarray}
	[\hat{\mathbb {H}},\sum\hat{\mathbf {X}}] =  [\hat{\mathbb {H}}, \sum\hat{\mathbf {Y}}] =  0\,.
\end{eqnarray}
With the help of these commutators, we can express Hall conductivity as
\begin{eqnarray}\label{CNBS}
	\sigma_{xy}&=&   -\frac{1}{i {\cal S}}\frac{e^2}{\hbar}\,\sum_{k\ne 0}   \,\Big[  \frac{\epsilon_{ij}}{({\cal E}_k - {\cal E}_0)^2}  \langle 0| [\hat{\mathbb {H}},\sum\hat{\boldsymbol{\xi}}_i | k \rangle    \langle k | [\hat{\mathbb {H}}, \sum\hat{\boldsymbol{\xi}}_j] | 0 \rangle  \Big]_{A=0} \,
\nonumber\\
	&=&  \frac{1}{i {\cal S}}\frac{e^2}{\hbar}\,\sum_{k \ne 0}   \, \epsilon_{ij}\,\Big[  \langle 0| \sum\hat{\boldsymbol{\xi}}_i | k \rangle    \langle k |  \sum\hat{\boldsymbol{\xi}}_j | 0 \rangle  \Big]_{A=0}
\nonumber\\
	&=&  \frac{1}{i {\cal S}}\frac{e^2}{\hbar}\,  \,\Big[  \langle 0|  [\sum\hat{\boldsymbol{\xi}}_x,\sum\hat{\boldsymbol{\xi}}_y ]| 0 \rangle  \Big]_{A=0}
\nonumber\\
	&=& -\frac{e}{{ B\cal  S}} \langle 0| \hat{\mathbb {N}} | 0 \rangle \,,
\end{eqnarray}
where
\begin{eqnarray}
	\hat{\mathbb {N}}=\int d^2x\,a^{\dagger}(x)a(x)\,.
\end{eqnarray}

If we denote by $M$ the number of electron states in a fully occupied Landau level, then
\begin{eqnarray}\label{BSM}
	|{B}|{\cal S}=M\Phi_0=\frac{2\pi \hbar}{|e|} M\,,
\end{eqnarray}
where
$\Phi_0$ is the magnetic flux quantum. Then we have
\begin{eqnarray}\label{CHCIF}
	\sigma_{xy}&=&\frac{\langle 0| \hat{\mathbb {N}} | 0 \rangle}{M}\, \frac{e^2}{h}{\rm sign}({B})\nonumber
	\\
	&=&\langle 0| \hat{\boldsymbol{\nu}} | 0 \rangle \, \frac{e^2}{h}{\rm sign}({B})\,,
\end{eqnarray}
where $\langle 0| \hat{\boldsymbol{\nu}} | 0 \rangle$ is the expectation value of the filling fraction for the ground state. Again, the Coulomb interactions do not affect the steps up to Eq.(\ref{CNBS}). From Eq.(\ref{CHCIF}) we see that the Hall conductivity is proportional to the number of electrons in the ground state. In the absence of interactions this is  just the number of electrons with the energies below the chemical potential. In the presence of interactions the total energy of the given state is already not  given by the sum of the occupied one - electron stares. Therefore, the meaning of the chemical potential is not so transparent. Notice, that the above derivation does not depend on the details of the ground state. Therefore, it is valid for both integer quantum Hall effect and the fractional quantum Hall effect.

\rv{Above in Eq. (\ref{CNBS}) it is assumed that the energies of the excited states differ from the ground state energy. Otherwise, the singularities are present. Therefore, we assumed that the considered system is gapped. We cannot say definitely how our expressions are changed for the gapless systems. It is worth mentioning, however that in the 3D systems the (non - topological) QHE may appear in the gapless systems. The example is given by Weyl semimetals. In any case the more detailed analysis is needed to extend Eq. (\ref{CNBS}) to the gapless systems, which has to involve the precise expressions for the ground state wave function as well as the wave functions of excited states, and their energies.}

\rv{We also notice that the above derivation fits the known phenomenological schemes of the FQHE. In particular, the one with the Laughlin’s wave function as the ground state  gives the correct filling fraction. This wave function is known to reproduce many features of the real ground state that minimizes energy of the system of interacting fermions.   Also the pattern of composite fermions does not exclude the application of our results. The composite fermion theory proposes a heuristical description of the interacting fermion systems.  Within this pattern the appearance of the fractional filling fraction ratio is explained. This appearance is enough to apply Eq. (\ref{CHCIF}). }

\section{Conclusions}
\label{concl}

To conclude, in the present paper we consider the quantum Hall effect in the systems with pair interactions between the electrons. We do not consider the influence of disorder on Hall conductivity (for the discussion of this issue see, for example, \cite{Alt0,Alt1}). The key tool used in our consideration is the standard operator formalism of equilibrium quantum field theory. Several useful identities of this formalism are accumulated in Appendix. Those identities were used while dealing with the commutation relations between particle momenta and coordinates. The coordinates are separated to those responsible for the ``center of orbit" motion and the local motion within the orbits. This representation is an extension to many - body systems of an old approach of \cite{old} (see also references therein).

 The main advantage of our approach is that arbitrary pair interactions between the electrons are taken into account. We observe that they do not affect the basic commutation relations used to derive the final expression for the Hall conductivity. This expression is given by Eq. (\ref{CHCIF}). It is the filling fraction operator averaged over the ground state. In the present paper we do not discuss the possible values of this average, and the nature of the ground state. The effective theories of the FQHE \cite{BFr,W, FrK, FZ, Z, FrS, FrST} prompt that the expectation value of $\hat{\boldsymbol{\nu}}$ may take the standard filling fraction values given by certain rational numbers. The microscopic theory that explains the appearance of these numbers remains out of the scope of the present paper. \rv{Notice, that at the given value of chemical potential modifications of interactions may change the value of the filling fraction via modification of the ground state.} We also do not prove robustness of the obtained result with respect to the smooth modification of the system, which is to be the subject of a separate investigation.



{The authors are grateful to I.Fialkovsky, C.Zhang, and M.Suleymanov for useful discussions.}










\appendix

\section{Commutators in second quantization}

\label{Commutators in second quantization}

Let us first check the relation between the commutators of operators in second quantization and commutators in quantum mechanics.  We deal with the single-body operators
\begin{eqnarray}
	\hat{\mathbb {F}}:=\int d^2x  \,a^{\dagger}(x )\hat{F}(x )a(x )\,,\quad\quad
	\hat{\mathbb {G}}:=\int d^2x  \,a^{\dagger}(x )\hat{G}(x )a(x )\,.
\end{eqnarray}
and the two-body operators
\begin{eqnarray}
	\hat{\mathbb {K}}:=\int d^2x'\int d^2x''  \,a^{\dagger}(x')a^{\dagger}(x'')\hat{K}(x',x'')a(x'')a(x')\,.
\end{eqnarray}
In both cases the Kernels $\hat{G}, \hat{F}, \hat{K}$ act as the operators on  $a(x)$ considered as functions of $x$.
First we prove Eq. (\ref{FG}):
\begin{eqnarray}
	[\hat{\mathbb {F}}, \hat{\mathbb {G}}]&=&\int d^2x\int d^2x'\,[a^{\dagger}(x )\hat{F}(x )a(x ),a^{\dagger}(x')\hat{G}(x')a(x')]\nonumber\\
	&=&\int d^2x\,a^{\dagger}(x )[\hat{F}(x ),\hat{G}(x)]a(x)
	\,,
\end{eqnarray}
We need the following commutators of creation and annihilation operators:
\begin{eqnarray}
	\{a^{\dagger}(x ),a(x')\}=\delta(x-x')\,, \quad
	\{a(x ),a(x')\}=\{a^{\dagger}(x ),a^{\dagger}(x')\}=0\,.
\end{eqnarray}
Then
\begin{eqnarray}
	[\hat{\mathbb {F}}, \hat{\mathbb {G}}]&=&\int d^2x\int d^2x'\,\Big(a^{\dagger}(x )\hat{F}(x )a(x )a^{\dagger}(x')\hat{G}(x')a(x')-a^{\dagger}(x')\hat{G}(x')a(x')a^{\dagger}(x )\hat{F}(x )a(x )\Big)\nonumber\\
	&=&\int d^2x\int d^2x'\,\Big(a^{\dagger}(x )\hat{F}(x )a(x )a^{\dagger}(x')\hat{G}(x')a(x')+a^{\dagger}(x )\overleftarrow{\hat{F}}(x )a^{\dagger}(x')a(x )\overrightarrow{\hat{G}}(x')a(x')
\nonumber\\
	&&-(a^{\dagger}(x')\hat{G}(x')a(x')a^{\dagger}(x )\hat{F}(x )a(x )+a^{\dagger}(x )\overleftarrow{\hat{F}}(x )a^{\dagger}(x')a(x )\overrightarrow{\hat{G}}(x')a(x'))\Big)\nonumber\\
	&=&\int d^2x\int d^2x'\,\Big(a^{\dagger}(x )\hat{F}(x )a(x )a^{\dagger}(x')\hat{G}(x')a(x')+a^{\dagger}(x )\overleftarrow{\hat{F}}(x )a^{\dagger}(x')a(x )\overrightarrow{\hat{G}}(x')a(x')
\nonumber\\
	&&-(a^{\dagger}(x')\hat{G}(x')a(x')a^{\dagger}(x )\hat{F}(x )a(x )+a^{\dagger}(x')a^{\dagger}(x )\overleftarrow{\hat{F}}(x )\overrightarrow{\hat{G}}(x')a(x')a(x ))\Big)\nonumber\\
	&=&\int d^2x\int d^2x'\,\Big(a^{\dagger}(x )\hat{F}(x )a(x )a^{\dagger}(x')\hat{G}(x')a(x')+a^{\dagger}(x )\overleftarrow{\hat{F}}(x )a^{\dagger}(x')a(x )\overrightarrow{\hat{G}}(x')a(x')
\nonumber\\
	&&-(a^{\dagger}(x')\hat{G}(x')a(x')a^{\dagger}(x )\hat{F}(x )a(x )+a^{\dagger}(x')\overleftarrow{\hat{G}}(x')a^{\dagger}(x )a(x')\overrightarrow{\hat{F}}(x )a(x ))\Big)\nonumber\\
	&=&\int d^2x\int d^2x'\,\Big(a^{\dagger}(x )\hat{F}(x )\hat{G}(x')a(x')\delta(x-x')+
\nonumber\\
	&&-a^{\dagger}(x')\hat{G}(x')\hat{F}(x )a(x )\delta(x-x')\Big)\nonumber\\
	&=&\int d^2x\,a^{\dagger}(x )[\hat{F}(x ),\hat{G}(x)]a(x)
	\,.
\end{eqnarray}
Next we prove Eq. (\ref{FK}):
\begin{eqnarray}
	[\hat{\mathbb {F}}, \hat{\mathbb {K}}]&=&\int d^2x\int d^2x'\int d^2x''\,[a^{\dagger}(x )\hat{F}(x )a(x ),a^{\dagger}(x')a^{\dagger}(x'')\hat{K}(x',x'')a(x'')a(x')]\nonumber\\
	&=&\int d^2x\int d^2x'\,a^{\dagger}(x )a^{\dagger}(x')[\hat{F}(x ),\hat{K}(x,x')+\hat{K}(x',x)]a(x')a(x)
	\,.
\end{eqnarray}
Note that
\begin{eqnarray}\label{aadad}
	[a(x),a^{\dagger}(x')a^{\dagger}(x'')]&=&a(x)a^{\dagger}(x')a^{\dagger}(x'')-a^{\dagger}(x')a^{\dagger}(x'')a(x)\nonumber\\
	&=&a(x)a^{\dagger}(x')a^{\dagger}(x'')+a^{\dagger}(x')a(x)a^{\dagger}(x'')-(a^{\dagger}(x')a^{\dagger}(x'')a(x)+a^{\dagger}(x')a(x)a^{\dagger})\nonumber\\
	&=&\{a(x),a^{\dagger}(x')\}a^{\dagger}(x'')-(a^{\dagger}(x')\{a^{\dagger}(x''),a(x)\}\nonumber\\
	&=&a^{\dagger}(x'')\delta(x-x'')-a^{\dagger}(x')\delta(x-x')
	\,.
\end{eqnarray}
Then the proof goes as
\begin{eqnarray}
	[\hat{\mathbb {F}}, \hat{\mathbb {K}}]&=&\int d^2x\int d^2x'\int d^2x''\,\Big(a^{\dagger}(x )\hat{F}(x )a(x )a^{\dagger}(x')a^{\dagger}(x'')\hat{K}(x',x'')a(x'')a(x')\nonumber\\
	&&-a^{\dagger}(x')a^{\dagger}(x'')\hat{K}(x',x'')a(x'')a(x')a^{\dagger}(x )\hat{F}(x )a(x )\Big)\nonumber\\
	&=&\int d^2x\int d^2x'\int d^2x''\,\Big(a^{\dagger}(x )\hat{F}(x )[a(x ),a^{\dagger}(x')a^{\dagger}(x'')]\hat{K}(x',x'')a(x'')a(x')\nonumber\\
	&&-\Big(a^{\dagger}(x')a^{\dagger}(x'')\hat{K}(x',x'')a(x'')a(x')a^{\dagger}(x )\hat{F}(x )a(x )-a^{\dagger}(x )\hat{F}(x )a^{\dagger}(x')a^{\dagger}(x'')a(x )\hat{K}(x',x'')a(x'')a(x')\Big)\Big)\nonumber\\
	&=&\int d^2x\int d^2x'\int d^2x''\,\Big(a^{\dagger}(x )\hat{F}(x )[a(x ),a^{\dagger}(x')a^{\dagger}(x'')]\hat{K}(x',x'')a(x'')a(x')\nonumber\\
	&&-\Big(a^{\dagger}(x')a^{\dagger}(x'')\hat{K}(x',x'')a(x'')a(x')a^{\dagger}(x )\hat{F}(x )a(x )-a^{\dagger}(x')a^{\dagger}(x'')a^{\dagger}(x )\overleftarrow{\hat{F}}(x )\overrightarrow{\hat{K}}(x',x'')a(x'')a(x')a(x )\Big)\Big)\nonumber\\
	&=&\int d^2x\int d^2x'\int d^2x''\,\Big(a^{\dagger}(x )\hat{F}(x )[a(x ),a^{\dagger}(x')a^{\dagger}(x'')]\hat{K}(x',x'')a(x'')a(x')\nonumber\\
	&&-(a^{\dagger}(x')a^{\dagger}(x'')\hat{K}(x',x'')[a(x'')a(x'),a^{\dagger}(x )]\hat{F}(x )a(x ))\Big)\nonumber\,.
\end{eqnarray}
Then substituting Eq. (\ref{aadad}) into the last step, we get
\begin{eqnarray}
	[\hat{\mathbb {F}}, \hat{\mathbb {K}}]&=&
	\int d^2x\int d^2x'\,\Big(a^{\dagger}(x)a^{\dagger}(x')\hat{F}(x )(\hat{K}(x,x')+\hat{K}(x',x))a(x')a(x)\nonumber\\
	&&-(a^{\dagger}(x)a^{\dagger}(x')(\hat{K}(x,x')+\hat{K}(x',x))\hat{F}(x )a(x')a(x))\Big)\nonumber\\
	&=&
	\int d^2x\int d^2x'\,\Big(a^{\dagger}(x)a^{\dagger}(x')[\hat{F}(x ),\hat{K}(x,x')+\hat{K}(x',x)]a(x')a(x)\Big)
	\,.
\end{eqnarray}


\end{document}